\documentclass{article}
\usepackage{amssymb}
\usepackage{amsmath}

\input{tcilatex}
\begin{document}

\begin{center}
\bigskip

\bigskip

\bigskip

\textbf{UPPER LIMIT ON THE SIZE OF GALACTIC HALO VIA HAMILTONIAN APPROACH}

\bigskip

\textbf{Amrita Bhattacharya,}$^{1,2,3,a}$\textbf{\ Ruslan Isaev,}$^{4,b}$ 
\textbf{K.B. Vijayakumar,}$^{5,c}$\textbf{and Kamal K. Nandi}$^{1,4,6,d}$

$\bigskip $

$\bigskip $

$^{1}$Department of Mathematics, University of North Bengal, Siliguri 734
013, India

$^{2}$Dipartimento di Matematica, Istituto \textquotedblleft
G.Castelnuovo\textquotedblright , Universit\`{a} di La Sapienza, P.le Aldo
Moro, 2, Rome, Italy

$^{3}$DiFarma, Universit\`{a} di Salerno, Via Ponte Don Melillo, 84084
Fisciano, Salerno, Italy

$^{4}$Joint Research Laboratory, Bashkir State Pedagogical University, Ufa
450000, Russia

$^{5}$Department of Physics, University of Mangalore, Mangalore 574 199,
India

$^{6}$Department of Theoretical Physics, Sterlitamak State Pedagogical
Academy, Sterlitamak 453103, Russia

\bigskip

$^{a}$Email: amrita\_852003@yahoo.co.in

$^{b}$Email: subfear@gmail.com

$^{c}$Email: kbvijayakumar@yahoo.com

$^{d}$E-mail: kamalnandi1952@yahoo.co.in
\end{center}

\bigskip

PACS number(s): 04.50.1h, 04.20.Cv

\begin{center}
\bigskip
\end{center}

\bigskip

\begin{center}
\textbf{Abstract}
\end{center}

Using the approach of autonomous Hamiltonian dynamical system, we attempt to
estimate the as yet unknown upper limit on the size of the galactic halo
surrounding galaxies (lenses). The key to determine the size of the halo is
to determine the maximum radius up to which stable material circular orbits
are admissible. We shall illustrate the approach by considering a solution
of the Weyl gravity containing a halo parameter $\gamma $. The upper limit
for several observed lenses are calculated for a typical value of $\gamma $
for definiteness, with and without the cosmological constant $\Lambda $.
These lenses (all having Einstein radius $R_{\text{E}}\approx 10^{23}$ cm)
consistently yield an upper limit $R_{\text{max}}^{\text{stable}}$($\simeq
4.25\times 10^{27}$ cm) inside the de Sitter radius only when $\gamma $ is 
\textit{negative}, while a positive $\gamma $ yields $R_{\text{max}}^{\text{%
stable}}$ always exceeding the de Sitter radius.

\begin{center}
----------------------------------------------------------
\end{center}

The issue of dark matter (halo matter), arising out of reconciling known
gravitational laws with observed flat rotation curves, is a challenging
problem in modern astrophysics. Observationally, so far, mass of the dark
matter appears to increase with distance in galaxies, but in clusters
exactly the reverse is true, the dark matter distribution actually decreases
with distance. Indeed, for certain dwarfs (such as DD0154) the rotation
curve has been measured to almost 15 optical length scales indicating that
the dark matter surrounding this object is extremely spread out (see, for a
review, Sahni [1]). The total mass of an individual galaxy is still somewhat
of an unknown quantity since a turn around to the $v\varpropto r^{-1/2}$ law
at large radii has not been convincingly observed. On the other hand, dark
matter is attractive and localized on scales less than the cosmological
distances where repulsive dark energy prevails. Therefore, one would like to
know if there could be any upper limit on the size of the dark matter halo.
The motivation here is to investigate this question using a certain model
theory.

Although classical Einstein's general relativity theory has been nicely
confirmed within the weak field regime of solar gravity and binary pulsars,
observations of flat rotation curves in the galactic halo still lack a
universally accepted satisfactory explanation. The most widely accepted
explanation hypothesizes that almost every galaxy hosts a large amount of
nonluminous matter, the so called gravitational dark matter, consisting of
unknown particles not included in the particle standard model, forming a
halo around the galaxy. One of the possibility could be that these particles
(WIMPs) encircling the galactic center are localized in a thick shell
providing the needed gravitational field [2]. The exact nature of either the
dark matter or dark energy is yet far too unknown beyond such speculations.
There also exist alternative theories, such as Modified Newtonian Dynamics
(MOND) [3,4], braneworld model [5], scalar field model [6] etc that attempt
to explain dark matter without hypothesizing them. A prominent model theory
is Weyl conformal gravity and a particular solution in the theory is the
Mannheim-Kazanas-de Sitter (MKdS) metric [7] that we shall consider here.

The MKdS solution contains two arbitrary parameters $\gamma $ and $k$ ($%
=\Lambda /3$) that are expected to play prominent roles on the galactic halo
and cosmological scales respectively. While the value of the cosmological
constant $\Lambda =1.29\times 10^{-56}$ cm$^{-2}$ is well accepted, there is
some ambiguity about the sign and magnitude of $\gamma $. From the flat
rotation curve data, Mannheim and Kazanas fix it to be positive and being of
the order of the inverse Hubble length, while Pireaux [8] argues for $%
\left\vert \gamma \right\vert \sim 10^{-33}$ cm$^{-1}$. Edery and Paranjape
[9] obtained a negative value from the gravitational time delay by galactic
clusters while the magnitude is still of the order of inverse Hubble length.

In the present context, we recall that massive neutral hydrogen atoms are
executing circular motions in the halo around the galactic center. The
redshifted light from those atoms are measured to determine their tangential
velocities [10,11]. Therefore it is necessary to consider massive test
particle orbits. However, due to conformal invariance of the theory,
geodesics for massive particles would in general depend on the conformal
factor $\Omega ^{2}(x)$, but here we assume a fixed conformal frame and not
considering other conformal variants of the metric.

The purpose of this article is to determine the maximum radius within which
there can exist stable circular orbits of massive test particles, beyond
which the orbits become unstable. The criterion of stability has been
originally suggested by Edery and Paranjape [9] because it provides a way to
the determination of a natural length scale or region of influence of
localized sources in the cosmological setting. The strategy we adopt here is
to frame the geodesic equation in the MKdS solution as a Hamiltonian system,
and based on it, analyze the stability, which seems to favor a negative $%
\gamma $.

The MKdS metric is given by [7,9] (vacuum speed of light, $c_{0}=1$):%
\begin{equation}
d\tau ^{2}=B(r)dt^{2}-\frac{1}{B(r)}dr^{2}-r^{2}(d\theta ^{2}+\sin
^{2}\theta d\varphi ^{2}),\text{ \ }B(r)=1-\frac{2M}{r}+\gamma r-kr^{2},%
\text{\ }
\end{equation}%
where $k$ and $\gamma $ are constants.

Using $u=1/r$, we get the following path equation for a test particle of
mass $m_{0}$ on the equatorial plane $\theta =\pi /2$:%
\begin{equation}
\frac{d^{2}u}{d\varphi ^{2}}=-u+3Mu^{2}-\frac{\gamma }{2}+\frac{M}{h^{2}}+%
\frac{1}{2h^{2}u^{2}}\left( \gamma -\frac{2k}{u}\right) ,
\end{equation}%
where $h=\frac{U_{3}}{m_{0}}$, the angular momentum per unit test mass. For
photon, $m_{0}=0\Rightarrow h\rightarrow \infty $ and one ends up with the
conformally invariant equation but without $k$ making its appearance:%
\begin{equation}
\frac{d^{2}u}{d\varphi ^{2}}=-u+3Mu^{2}-\frac{\gamma }{2}.
\end{equation}%
In the Schwarzschild-de Sitter (SdS) metric, such a disappearance has been
noted for long [12] but here we find that it occurs despite the presence of $%
\gamma $ in the metric. The impact of $k$ and $\gamma $ on light bending has
been investigated elsewhere, in Refs.[13-16].

Analysis of dynamical system involves converting the second order equation
into two first order equations [17,18]. For this purpose, we introduce the
notation%
\begin{equation}
u=x\text{, }y=\overset{.}{x}=\frac{dx}{d\varphi }
\end{equation}%
to reduce Eq.(2) into a pair of first order autonomous system in the ($x,y$)
phase plane%
\begin{align}
\overset{.}{x}& =X(x,y)=y \\
\overset{.}{y}& =Y(x,y)=a+bx+cx^{2}+dx^{-2}+ex^{-3}
\end{align}%
where%
\begin{equation}
a=\frac{M}{h^{2}}-\frac{\gamma }{2}\text{, }b=-1\text{, }c=3M\text{, }d=%
\frac{\gamma }{2h^{2}}\text{, }e=-\frac{k}{h^{2}}.
\end{equation}

First we discuss stability of circular orbits of light, though their
stability is not essentially needed. Recall that even in the Schwarzschild
spacetime, circular light orbits at $R=3M$ are unstable. Nonetheless, it is
instructive to have a look at this aspect in the MKdS solution.

\textit{(a) Massless particle motion}

Light motion occurs in circular orbits defined by $R(2+\gamma R)-6M=0$ [see
Eq.(11) below] because $h^{2}\rightarrow \infty $, which implies that $d=e=0$
but $\gamma \neq 0$. The equilibrium points are given by $\overset{.}{x}=0$, 
$\overset{.}{y}=0$, which yield $\left( \frac{-b+\sqrt{b^{2}-4ac}}{2c}%
,0\right) $ and $\left( \frac{-b-\sqrt{b^{2}-4ac}}{2c},0\right) .$ To locate
these points on the real phase plane ($x,y$), we must have $\alpha
^{2}\equiv b^{2}-4ac=1+6\gamma M\geq 0.$ Now $\alpha ^{2}=0\Rightarrow
\gamma =-\frac{1}{6M}$, so the equilibrium points reduce to one single point
given by $P:\left( \frac{1}{6M},0\right) $. For\ $\alpha ^{2}>0$, or $\gamma
>-\frac{1}{6M}$, there are two distinct equilibrium points points $Q_{\pm
}:\left( \frac{1+\alpha }{6M},0\right) $ where $\ \alpha =$ $\pm \sqrt{%
1+6\gamma M}$. Thus $Q_{\pm }$ correspond to two $\gamma -$dependent light
radii $R_{\pm }=\frac{6M}{1\pm \sqrt{1+6\gamma M}}$, which expand as follows%
\begin{equation}
R_{+}=\frac{-1+\sqrt{1+6M\gamma }}{\gamma }\approx 3M+O(\gamma ),
\end{equation}%
\begin{equation}
R_{-}=\frac{-1-\sqrt{1+6M\gamma }}{\gamma }\approx -3M-\frac{2}{\gamma }%
+O(\gamma ).
\end{equation}%
We have from Eq.(23) below%
\begin{equation}
q_{0\pm }=1-\frac{6M}{R_{\pm }}
\end{equation}%
which yields $q_{0+}=-\sqrt{1+6M\gamma }<0$ leading to unstable radius at $%
R=R_{+}$, while $q_{0-}=\sqrt{1+6M\gamma }>0$ showing that $R=R_{-}$ is a
stable radius. The basic constraint (reality condition) for both is that $%
\gamma >-\frac{1}{6M}$. With, say, $\gamma =-7\times 10^{-28}$ cm$^{-1}$
(value inspired by Ref.[9]), the constraint is always satisfied for known
lenses (say, for $M=2.9\times 10^{18}$cm, Abell 2744). From Eq.(9), then we
get the value $R_{-}=2.86\times 10^{27}$ cm, at which there is stability.
However, stability of massive particle circular orbits is more relevant,
which we examine next.

\textit{(b) Massive particle motion }

The equilibrium points are given by $\overset{.}{x}=0$ and $\overset{.}{y}=0$%
. The equation $\overset{.}{x}=0$ gives $r=R=$ constant, while $\overset{.}{y%
}=0$ gives%
\begin{equation}
h^{2}=-\frac{2MR^{2}+R^{4}(\gamma -2kR)}{R(2+\gamma R)-6M}.
\end{equation}%
The autonomous system (5), (6) can be phrased as a Hamiltonian system as
follows 
\begin{align}
\frac{\partial H}{\partial x}& =-Y\left( x,y\right)
=-(a+bx+cx^{2}+dx^{-2}+ex^{-3}) \\
\frac{\partial H}{\partial y}& =X\left( x,y\right) =y.
\end{align}%
The necessary and sufficient condition for the system (12),(13) to be a
Hamiltonian system, namely, $\frac{\partial X}{\partial x}+\frac{\partial Y}{%
\partial y}=0,$ is fulfilled for all $x$ and $y$. Moreover, $\frac{dH}{%
d\varphi }=0$ and therefore $H\left( x,y\right) =$ constant (independent of $%
\varphi $). Integrating Eqs.(12),(13), we get%
\begin{equation}
H(x,y)=-(ax+\frac{b}{2}x^{2}+\frac{c}{3}x^{3}-dx^{-1}-\frac{e}{2}x^{-2})+u(y)
\end{equation}%
\begin{equation}
H(x,y)=\frac{1}{2}y^{2}+v(x)
\end{equation}%
where $u(y)$ and $v(x)$ are arbitrary functions subject to the consistency
of Eqs.(12) and (13). These two equations will match only if 
\begin{equation}
u(y)=\frac{1}{2}y^{2}+C
\end{equation}%
\begin{equation}
v(x)=-(ax+\frac{b}{2}x^{2}+\frac{c}{3}x^{3}-dx^{-1}-\frac{e}{2}x^{-2})+E
\end{equation}%
where $C$, $E$ are arbitrary constants. The family of Hamiltonian paths on
the phase plane are given by%
\begin{equation}
H(x,y)=\frac{1}{2}y^{2}-(ax+\frac{b}{2}x^{2}+\frac{c}{3}x^{3}-dx^{-1}-\frac{e%
}{2}x^{-2})+G
\end{equation}%
where $G$ is a parameter. It follows that 
\begin{equation}
\frac{\partial ^{2}H}{\partial x^{2}}=-(b+2cx-2dx^{-3}-3ex^{-4})
\end{equation}%
\begin{equation}
\frac{\partial ^{2}H}{\partial y^{2}}=1
\end{equation}%
\begin{equation}
\frac{\partial ^{2}H}{\partial x\partial y}=0.
\end{equation}%
As before, the equilibrium points occur when $X=0$ and $Y=0$, which give the
values of the orbit radius $r=R=$ constant and $h^{2}$ as in Eq.(11). The
quantity determining stability is [17] 
\begin{equation}
q\equiv \frac{\partial ^{2}H}{\partial x^{2}}\frac{\partial ^{2}H}{\partial
y^{2}}-\left( \frac{\partial ^{2}H}{\partial x\partial y}\right) ^{2}.
\end{equation}%
Putting the value of $h^{2}$, we get at the equilibrium points, suffixed by
zero, the following expression%
\begin{equation}
q_{0}=1-\frac{6M}{R}+\frac{R(3kR-\gamma )[R(2+\gamma R)-6M]}{%
R^{2}(2kR-\gamma )-2M}.
\end{equation}%
When $q_{0}>0$ at any point $P:(R,0)$, we say that $P$ is a stable center,
but it is an unstable saddle point if $q_{0}<0$. When $q_{0}=0$, it is an
inflection point where the system begins to become unstable. Thus $q_{0}=0$
gives $R=R_{\text{max}}^{\text{stable}}$, beyond which the orbits begin to
become unstable. There is also a singular radius $R=R_{\text{sing}}$ where $%
q_{0}$ blows up. In all cases we have studied, $R_{\text{sing}}<R_{\text{E}}$
for $\gamma $ negative and $R_{\text{sing}}$ $>R_{\text{dS}}$ for $\gamma $
positive. In either case, the presence of singularity is of no concern.

We shall calculate the values of $q_{0}$ for observed lenses. Fortunately,
due to advances in technology, mass $M$ and Einstein radius $R_{\text{E}}$
for several lenses are available. The observed lens masses $M$ are used to
compute the values of $q_{0}$ marching from the Einstein radius $R_{\text{E}%
} $ out to $R_{\text{dS}}$. A meaningful stable radius should lie between $%
R_{\text{E}}$ and $R_{\text{dS}}=\sqrt{\Lambda /3}=1.52\times 10^{28}$cm.
Fig.1 illustrates for one of the observed lenses (Abell 2744, $M=2.90\times
10^{18} $cm, $R_{\text{E}}=2.97\times 10^{23}$cm), which shows that stable
material radii exist for \textit{all} radii $R\geq R_{\text{E}}$ when $%
\gamma =7\times 10^{-28}$ cm$^{-1}$ suggesting that the halo might extend
even beyond the dS radius, which is improbable. Also, the singular radius
occurs at $R_{\text{sing}}=8.14\times 10^{28}$cm $>R_{\text{dS}}$. (Fig. 2).
On the other hand, when $\gamma =-7\times 10^{-28}$ cm$^{-1}$, stability is
achieved up to a radius $R=4.25\times 10^{27}$ cm $<R_{\text{dS}}$ beyond
which instability begins. In this case, the singular radius occurs at $R_{%
\text{sing}}=9.11\times 10^{22}$cm $<R_{\text{E}}$ (Fig. 3). Thus the
maximum radius up to which stable material circular orbits are admissible is 
$R_{\text{max}}^{\text{stable}}=4.25\times 10^{27}$ cm. This limit is not
much sensitive either to the value of $\Lambda $ or to slight magnitude
variations in $\gamma $. It can be verified that, for other lens data (as
listed in Ishak et al. [14], but not tabulated here), the upper limit
remains very nearly the same.

We conclude the following: A finite upper limit $R_{\text{max}}^{\text{stable%
}}$ on the halo radius exists within $R_{\text{dS}}$ only for a negative $%
\gamma $. We are aware that the tool used here is probably far too
simplistic than the ground situation, but it is nevertheless a good one,
which allows us to qualitatively infer a negative $\gamma $. Much less is
yet conclusively known about the halo content or size, which engenders many
theoretical models, not to mention stiff observational challenges. We
considered here the MKdS solution by way of illustration but any other
metric solution can be used likewise.

One of us (AB) wishes to thank the authorities of the Universit\`{a}\ di "La
Sapienza", Rome, Italy for financial support during the work. All authors
thank Guzel N. Kutdusova and Sonali Sarkar for technical assistance.

\textbf{Figure captions}

Fig.1. The upper line corresponds to $\gamma =7\times 10^{-28}$ cm$^{-1}$
and the lower line to $\gamma =-7\times 10^{-28}$ cm$^{-1}$ in the case of
Abell 2744. The plot remains essentially the same for other lenses tabulated
in Ref.[14].

Fig. 2. Curves I correspond to $\gamma =7\times 10^{-28}$ cm$^{-1}$ giving a
singularity of $q_{0}$ at $R_{\text{sing}}=8.14\times 10^{28}$cm $>R_{\text{%
dS}}$.

Fig. 3. Curves II correspond to $\gamma =-7\times 10^{-28}$ cm$^{-1}$ giving
a singularity of $q_{0}$ at $R_{\text{sing}}=9.11\times 10^{22}$cm $<R_{%
\text{E}}$.

\textbf{References}

[1] V. Sahni, \textquotedblleft Dark matter and dark
energy\textquotedblright\ [arXiv: astro-ph/0403324 v3]

[2] K. Lake, \textquotedblleft Galactic halos are Einstein clusters of
WIMPs\textquotedblright\ [arXiv: gr-qc/0607057 v3]

[3] J. D. Bekenstein and R. H. Sanders, Astrophys. J. \textbf{429}, 480
(1994), J. D. Bekenstein, Phys. Rev. D \textbf{70}, 083509 (2004); Erratum: 
\textit{ibid}. D \textbf{71}, 069901 (2005)

[4] M. Milgrom, Astrophys. J. \textbf{270}, 365 (1983); \textit{ibid.} 
\textbf{270}, 371 (1983); \textit{ibid.} \textbf{270}, 384 (1983); Phys.
Rev.D \textbf{80}, 123536 (2009)

[5] K. K. Nandi, A.I. Filippov, F. Rahaman, Saibal Ray, A. A. Usmani, M.
Kalam and A. DeBenedictis, Mon. Not. R. Astron. Soc. \textbf{399}, 2079
(2009); F. Rahaman, M. Kalam, A. DeBenedictis, A.A. Usmani and Saibal Ray,
Mon. Not. R. Astron. Soc. \textbf{389}, 27 (2008)

[6] T. Matos, F.S. Guzm\'{a}n and D. Nu\~{n}ez, Phys. Rev. D \textbf{62},
061301 (2000); K.K. Nandi, I. Valitov, and N.G. Migranov, Phys. Rev. D 
\textbf{80}, 047301 (2009)

[7] P. D. Mannheim and D. Kazanas, Astrophys. J. \textbf{342}, 635 (1989);
P. D. Mannheim, Astrophys. J. \textbf{479}, 659 (1997); P. D. Mannheim,
Phys. Rev. D \textbf{75}, 124006 (2007). See also the review:\ P.D.
Mannheim, Prog. Part. and Nucl. Phys. \textbf{56}, 340 (2006)

[8] S. Pireaux, Class. Quant. Grav. \textbf{21}, 4317 (2004)

[9] A. Edery and M. B. Paranjape, Phys. Rev. D \textbf{58}, 024011 (1998)

[10] K. Lake, Phys. Rev. Lett. \textbf{92}, 051101 (2004)

[11] U. Nucamendi, M. Salgado and D. Sudarsky, Phys. Rev. D \textbf{63},
125016 (2001)

[12] J.N. Islam, Phys. Lett. A \textbf{97}, 239 (1983)

[13] W. Rindler and M. Ishak, Phys. Rev. D \textbf{76}, 043006 (2007)

[14] M. Ishak, W. Rindler, J. Dossett, J. Moldenhauer and C. Allen, Mon.
Not. Roy. Astron. Soc. \textbf{388}, 1279 (2008)

[15] Amrita Bhattacharya, Ruslan Isaev, Massimo Scalia, Carlo Cattani and
Kamal K. Nandi, \textquotedblleft Rindler-Ishak method: Light deflection in
Weyl gravity\textquotedblright\ [arXiv: gr-qc/1002.2601]

[16] Amrita Bhattacharya, Guzel M. Garipova, Alexander A. Potapov, Arunava
Bhadra and Kamal K. Nandi, \textquotedblleft The vacuole model revisited:
new repulsive terms in the second order deflection of
Light\textquotedblright\ [arXiv: gr-qc/0910.1112]

[17] D.W. Jordan and P. Smith, \textit{Nonlinear Ordinary Differential
Equations}, 3rd edition (Oxford University Press, Oxford, 1999)

[18] A. Palit, A. Panchenko, N.G. Migranov, A. Bhadra and K.K. Nandi, Int.
J. Theor. Phys. \textbf{48}, 1271 (2009)

\FRAME{ftbpF}{6.1134in}{3.9704in}{0in}{}{}{figure 1.eps}{\special{language
"Scientific Word";type "GRAPHIC";maintain-aspect-ratio TRUE;display
"USEDEF";valid_file "F";width 6.1134in;height 3.9704in;depth
0in;original-width 22.102in;original-height 14.3187in;cropleft "0";croptop
"1";cropright "1";cropbottom "0";filename 'Figure 1.eps';file-properties
"XNPEU";}}

Fig.1. The upper line corresponds to $\gamma =7\times 10^{-28}$ cm$^{-1}$
and the lower line to $\gamma =-7\times 10^{-28}$ cm$^{-1}$ in the case of
Abell 2744. The plot remains essentially the same for other lenses tabulated
in Ref.[14].

\FRAME{ftbpF}{6.8415in}{4.4296in}{0in}{}{}{figure 2.eps}{\special{language
"Scientific Word";type "GRAPHIC";maintain-aspect-ratio TRUE;display
"USEDEF";valid_file "F";width 6.8415in;height 4.4296in;depth
0in;original-width 6.7767in;original-height 4.3785in;cropleft "0";croptop
"1";cropright "1";cropbottom "0";filename 'Figure 2.eps';file-properties
"XNPEU";}}

Fig. 2. Curves I correspond to $\gamma =7\times 10^{-28}$ cm$^{-1}$ giving a
singularity of $q_{0}$ at $R_{\text{sing}}=8.14\times 10^{28}$cm $>R_{\text{%
dS}}$.

\FRAME{ftbpF}{7.7357in}{4.7098in}{0in}{}{}{figure 3.eps}{\special{language
"Scientific Word";type "GRAPHIC";maintain-aspect-ratio TRUE;display
"USEDEF";valid_file "F";width 7.7357in;height 4.7098in;depth
0in;original-width 7.6666in;original-height 4.6561in;cropleft "0";croptop
"1";cropright "1";cropbottom "0";filename 'Figure 3.eps';file-properties
"XNPEU";}}

Fig. 3. Curves II correspond to $\gamma =-7\times 10^{-28}$ cm$^{-1}$ giving
a singularity of $q_{0}$ at $R_{\text{sing}}=9.11\times 10^{22}$cm $<R_{%
\text{E}}$.

\end{document}